\documentclass[useAMS,usenatbib]{mn2e}

\usepackage{latexsym,amsmath,amstext} 
\usepackage[dvips]{graphicx}          
\usepackage{color}
\usepackage{graphicx}
\usepackage{natbib}
\bibpunct{(}{)}{;}{a}{}{,}

\title[TTVs of Laplace-resonant systems]
{Detection of Laplace-resonant three-planet systems from transit timing variations}

\author[A.-S. Libert and S. Renner]{A.-S. Libert$^{1,2,3}$\thanks{E-mail:
anne-sophie.libert@fundp.ac.be (ASL); stefan.renner@univ-lille1.fr (SR)} and S. Renner$^{2,3}$\\
$^{1}$ NaXys, Department of Mathematics, University of Namur, 8 Rempart de la Vierge, 5000 Namur, Belgium\\
$^{2}$ Laboratoire d'Astronomie de Lille, Universit\'e Lille 1, 1 impasse
de l'Observatoire, 59000 Lille, France \\ 
$^{3}$ Institut de M\'ecanique C\'eleste et de Calcul des Eph\'em\'erides, UMR 8028 du CNRS, 77 avenue Denfert-Rochereau, 75014 Paris, France}

\begin{document}

\pagerange{\pageref{firstpage}--\pageref{lastpage}} \pubyear{2012}

\maketitle

\label{firstpage}

%
%
%

\begin{abstract}
Transit timing variations (TTVs) are useful to constrain the existence of perturbing planets, especially in resonant systems where the variations are strongly enhanced. Here we focus on Laplace-resonant three-planet systems, and assume the inner planet transits the star. A dynamical study is performed for different masses of the three bodies, with a special attention to terrestrial planets. We consider a maximal time-span of $\sim 100$ years and discuss the shape of the inner planet TTVs curve. Using frequency analysis, we highlight the three periods related to the evolution of the system: two periods associated with the Laplace-resonant angle and the third one with the precession of the pericenters. These three periods are clearly detected in the TTVs of an inner giant planet perturbed by two terrestrial companions. Only two periods are detected for a Jupiter-Jupiter-Earth configuration (the ones associated with the giant interactions) or for three terrestrial planets (the Laplace periods). However, the latter system can be constrained from the inner planet TTVs. We finally remark that the TTVs of resonant three or two Jupiter systems mix up, when the period of the Laplace resonant angle matches the pericenter precession of the two-body configuration. This study highlights the importance of TTVs long-term observational programs for the detection of multiple-planet resonant systems.

\end{abstract}




\section{Introduction}

More than 100 multi-planetary systems have been detected\footnote{see the updated database exoplanet.eu (\citealt{Sch11})}. 
At present, at least three extrasolar systems are 
believed to be in a  multiply-resonant configuration, 
similar to the Laplace resonance of the Jupiter's Galilean moons: HR~8799 (e.g. \citealt{Rei09}) and Gliese 
876 (\citealt{Riv10}) in a Laplace 4:2:1 configuration, and the KOI-500 
five-body exosystem 
that presents two three-body resonances (\citealt{Lis11}).  

In formation theories, it has been shown that, at least in the case of low-mass planets such as the ones in our 
Solar System, gas-driven migration can force the planets to enter into a multiple-planet resonance (\citealt{Mor07}). 
Recently, \citet{Lib11a,Lib11b} have shown that a Laplace three-planet resonance can be achieved for Jupiter-mass planets 
and, provided that enough eccentricity damping is exerted on the planets, such a three-planet system could be stable 
after the dissipation of the gas disc.   

Multiple-planet resonances can act as a phase-protection mechanism, 
especially for systems consisting of a lot of massive bodies in a particularly compact 
configuration. In their study of Gliese 876, \citet{Riv10} have shown that the Laplace resonance is 
essential for the long-term stability of the system. A lot of KOIs also show evidence of their 
proximity to multiple-planet commensurabilities (\citealt{Lis11}), but it needs confirmation since the orbital parameters of 
the planets are still unknown. 

The TTVs method is a powerful technique to infer the existence of additional non-transiting planets 
(\citealt{Mir02,Ago05,Hol05}). Indeed a companion (not necessarily transiting) planet interacts 
gravitationally with the transiting planet and generates deviations of the transit times with respect to the strictly 
periodic Keplerian case. This is especially the case for resonant systems where the variations are strongly enhanced. 
The TTVs technique has been applied using ground-based data (e.g. \citealt{Ste05,Col07}), MOST data (e.g. \citealt{Mil08}), 
HST data (e.g. \citealt{AgoSt07}) and recently the space missions CoRoT (e.g. \citealt{Bea09,Csi10}) and 
Kepler (e.g. \citealt{Bor10,Bor11}). However, deriving the orbital elements and the mass 
of the perturber from TTVs is a difficult inverse problem (e.g. \citealt{Nes08}), and different 
configurations can generate similar TTVs with the same dominant period (e.g. \citealt{For07,Bou12}). 
Despite these difficulties, \citet{Nes12} have recently achieved the first detection and characterization 
of a non-transiting planet by TTVs for the KOI-872 system. The TTVs of other Kepler confirmed/candidate systems 
are currently under investigation (\citealt{For11,For12,Ste12,Fab12}). The TTVs method is also promising for 
detecting Trojan planets (e.g. \citealt{For07}) and moons of transiting planets (e.g. \citealt{Kip09}).

Here we propose an initial qualitative exploration of the theoretical TTVs of a three-planet multiply-resonant system, assuming that only the inner body transits the star. We analyze the shape of its TTVs curve on a maximal time-span of $\sim 100$ years, in order to identify relevant features of the resonant evolution of the whole system. This allows us to define a simple criterion to distinguish three-body resonant systems from two-body ones. We only consider here the Laplace 4:2:1 resonance, varying the mass values of the three planets, but our results could be transposed to another three-planet resonance.

The paper is organized as follows. In Section \ref{secTTVs}, we analyze the TTVs signal of a Laplace-resonant system composed of three giant planets, identifying the periods with the time variation of the orbital elements. Section \ref{secter} focuses on the possible detection of terrestrial planets in multiply-resonant systems. A discussion is given in Section~\ref{secdisc}, while our results are summarized in Section~\ref{ccl}.

 \begin{figure}
 \hspace{-8.8cm}
\rotatebox{270}{\includegraphics[width=12cm]{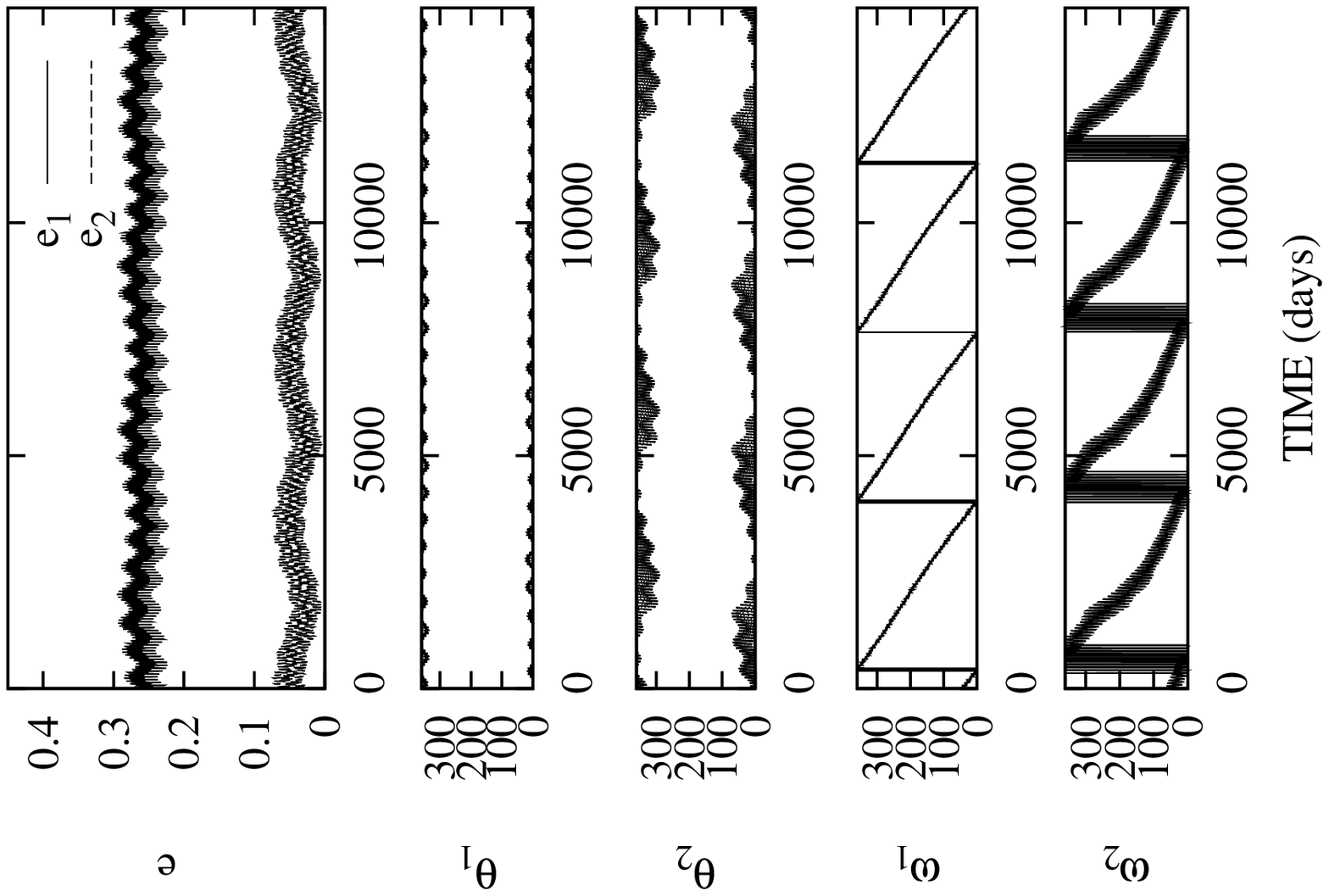}}
\rotatebox{270}{\includegraphics[height=8.3cm]{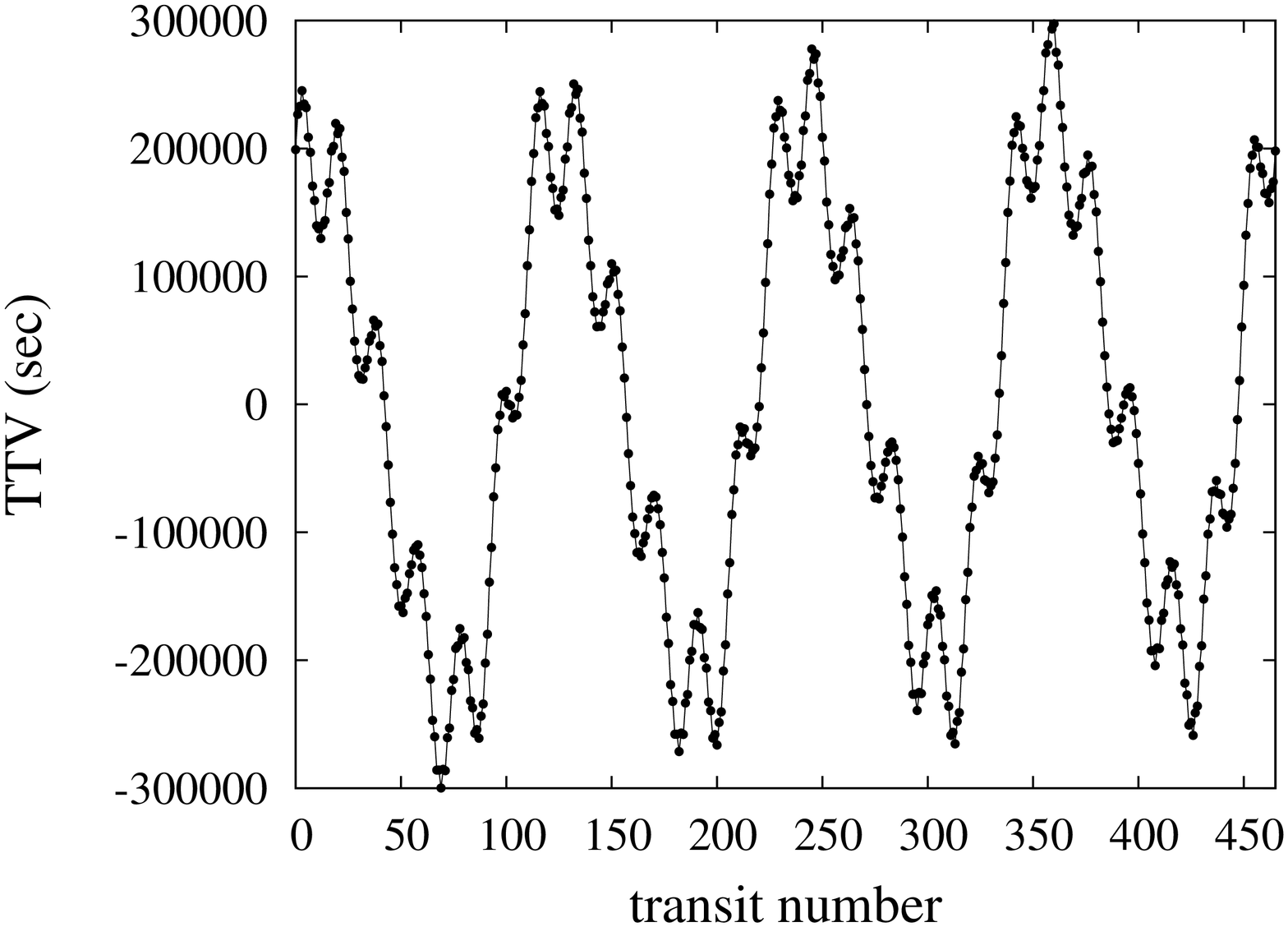}}
 \caption{Dynamical evolution of the two-planet resonant Gliese~876 system on $40$ years. 
 The parameters of planets b and c are issued from Table 2 of \citet{Riv10}. Masses of the planets 
 are $0.72$ and $2.27$ $M_{J}$, and the orbital periods are $30.1$ and $61.1$ days, respectively. 
 From top to bottom: eccentricities, resonant angles ($\theta_i=\lambda_1-2\lambda_2+\omega_i$, $i=1,2$), pericenter arguments, and TTVs of the inner planet.}
 \label{figlgterme_2pl}
 \end{figure}

 \begin{figure*}
 \centering
 \rotatebox{270}{\includegraphics[width=12cm]{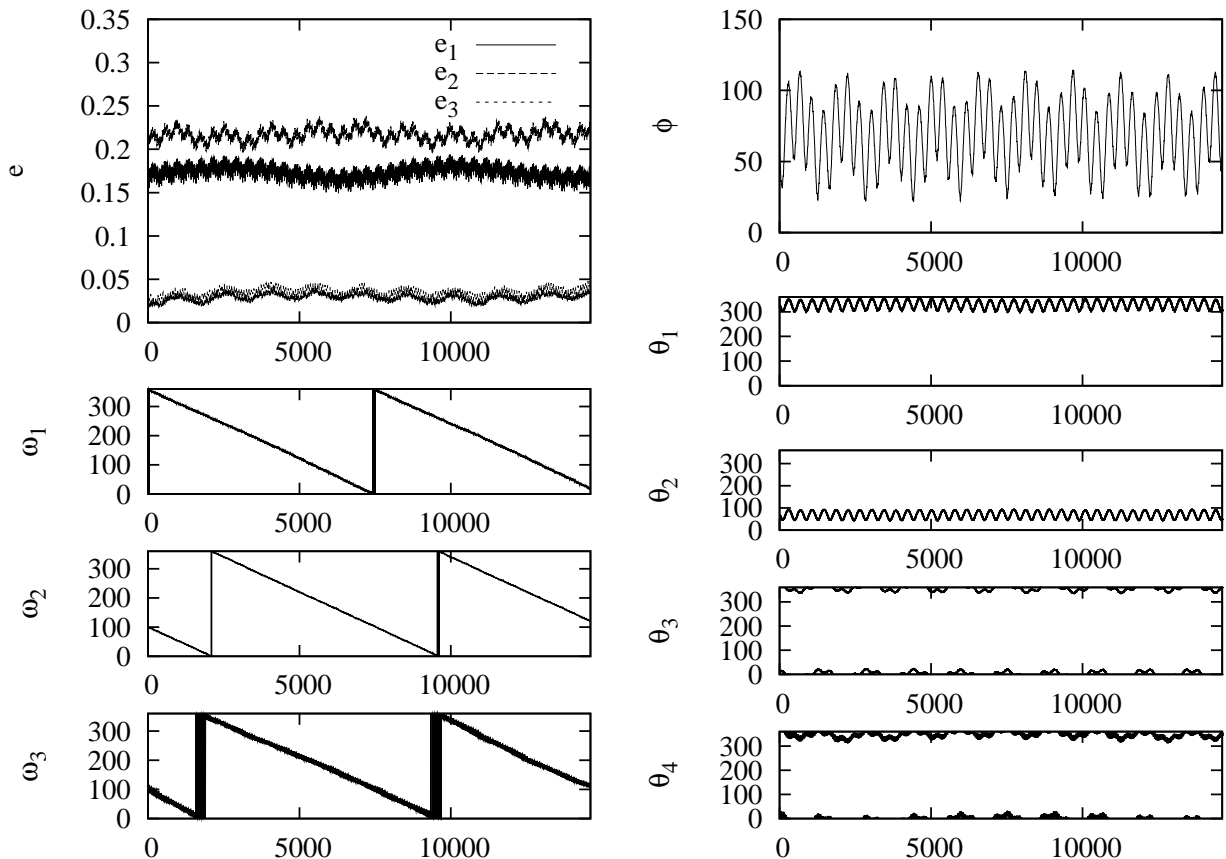}}
 \caption{Dynamical evolution (on $40$ years) of a Laplace-resonant three-planet system consisting 
 of three giant planets of $1.5 M_J$ with orbital periods of $20.1$, $40.5$ and $81.2$ days respectively. 
 The parameters are issued from the work of \citet{Lib11b}. Time is given in days.}
 \label{figlgterme_casdebase}
 \end{figure*}

 \begin{figure}
 \centering
 \rotatebox{270}{\includegraphics[width=6cm]{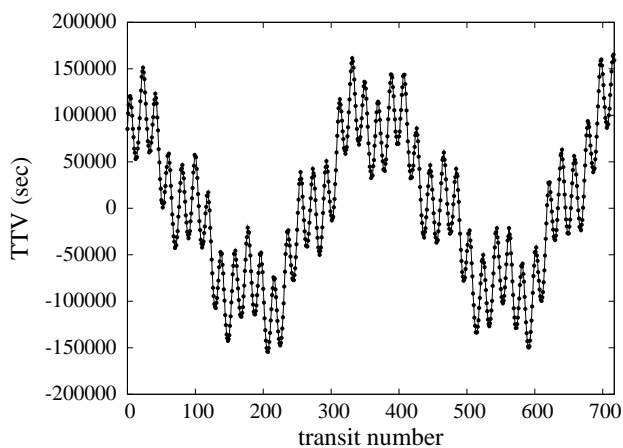}}
 \caption{TTVs of the inner planet of the Laplace-resonant three-planet system of Figure \ref{figlgterme_casdebase}. The time interval is $40$ years.}
 \label{TTVs_casdebase}
 \end{figure}

 \begin{table}
   \caption{Frequency analysis of the TTVs curve and the time variations of the 
   resonant angles ($\theta_i$), Laplace angle ($\phi$) and arguments of the pericenter ($\omega$). Top : the resonant two-planet Gliese 876 system, see Fig.~\ref{figlgterme_2pl}. Bottom : the Laplace-resonant three-planet system, see Fig.~\ref{figlgterme_casdebase}. The main periods are listed by decreasing amplitude of the trigonometric term ($c_1$ is the highest amplitude).}
     \begin{center}
       \begin{tabular}{r|cccc}
         \hline
         Periods (days) & TTVs &  $\theta_1$ & $\theta_2$ & $\omega$    \\
         \hline
         3\,559 & $c_1$ &  & $c_1$ & ${\bf c_1}$   \\
         589 & $c_2$ & $c_1$ &  &     \\
         \hline
       \end{tabular}
     \end{center}
     \begin{center}
       \begin{tabular}{r|ccccccc}
         \hline
         Periods (days) & TTVs &  $\theta_1$ & $\theta_2$ & $\theta_3$ &$\theta_4$ &$\phi$ & $\omega$    \\
         \hline
         7\,507 & $c_1$ & & &     & &    &  ${\bf c_1}$   \\
         391 & $c_2$ & $c_1$ & $c_1$ & &  & $c_1$ &  \\
	 1522 & $c_3$ & &  &  $c_1$ &  $c_1$ & $c_2$ &  \\
         \hline
       \end{tabular}
     \end{center}
   \label{tablefreq}
 \end{table}

 \begin{figure}
 \rotatebox{270}{\includegraphics[width=5.9cm]{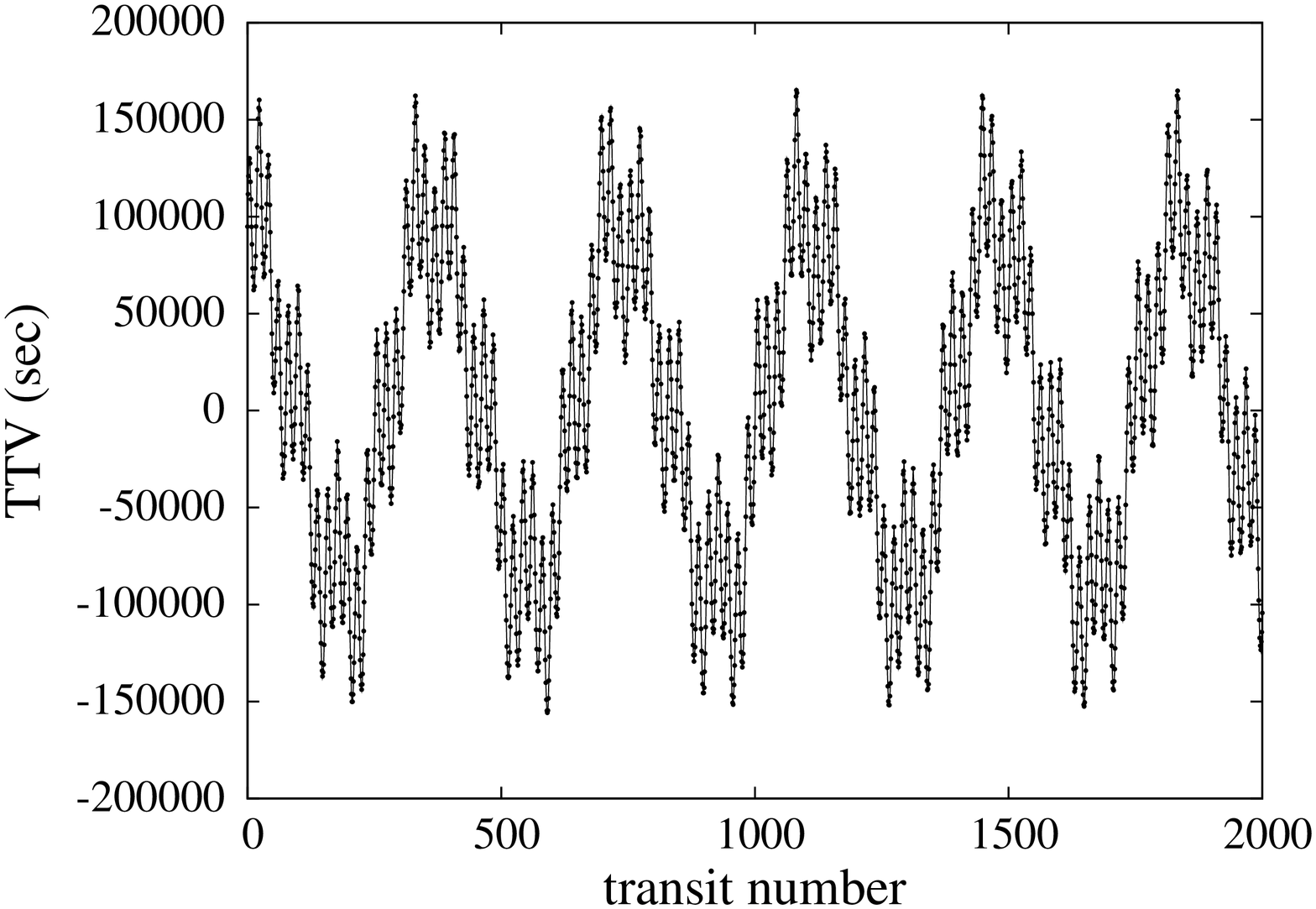}}
 \rotatebox{270}{\includegraphics[width=5.9cm]{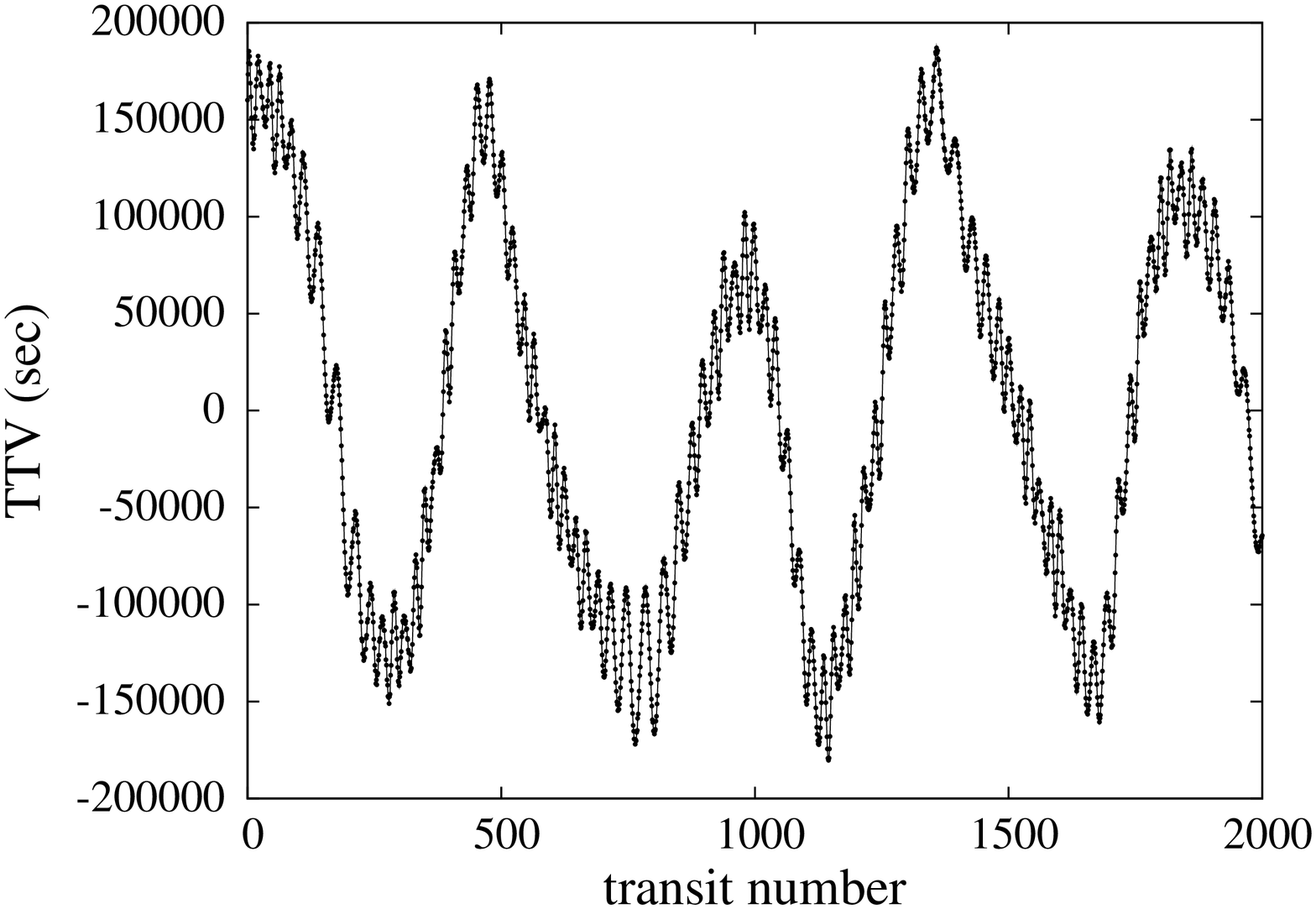}}
 \caption{Comparison of the TTVs of the inner planet for the system given in Figure \ref{figlgterme_casdebase} (top panel) and for a modified system where the outer planet has been removed from the Laplace resonance (bottom panel).}
 \label{fig_dege}
 \end{figure}

\section{TTVs signal of a Laplace-resonant system}\label{secTTVs}

We assume in this work that the inner planet of the system transits the star, such that our analysis relies on the TTVs curve of this planet only. We first recall the features of 
the TTVs signal of a two-planet resonant system.

An example of a two-planet resonant dynamics is shown in Figure \ref{figlgterme_2pl}, 
where the 2:1 mean-motion resonant evolution of the two-planet Gliese 876 system is given for 40 years (parameters of \citet{Riv10}). 
The orbital periods of the two planets are $30.1$ and $61.1$ days, respectively. As stated by \citet{Hol05} and \citet{Ago05}, 
the TTVs curve of the inner planet in Figure \ref{figlgterme_2pl} (bottom panel) shows two periods: a short one associated with the resonant angles ($\theta_i=\lambda_1-2\lambda_2+\omega_i$, $i=1,2$, $\lambda$ being the mean 
longitude and $\omega$ the argument of the pericenter), while the longer one corresponds 
to the precession rate of the arguments of the pericenters. Their values are given in 
Table~\ref{tablefreq} (top) where, by resorting to frequency analysis (\citealt{Las93}), 
we compute the main periods of the TTVs and the orbital elements.
 They are listed by decreasing amplitude of the trigonometric term : the coefficient $c_1$ represents the highest amplitude, and bold type $c_1$ indicates the precession rate of an angular variable in circulation. Let us note that only the period associated with the trigonometric term of highest amplitude is indicated for 
  $\theta_i$ and $\omega$ (the other period and/or a period combination are also present, as can be seen in Fig.~\ref{figlgterme_2pl}).

Let us perform a similar study for a system of three planets in a Laplace resonance. We choose to 
analyze one of the Laplace-resonant co-planar systems formed by migration in a gaseous disc in the 
work of \citet{Lib11b}. It consists of three planets with equal masses of $1.5 M_J$ ($M_J=$ Jupiter's mass) and orbital 
periods of $20.1$, $40.5$ and $81.2$ days respectively (see Fig.~\ref{figlgterme_casdebase}). 
The TTVs of the inner planet are shown in Figure \ref{TTVs_casdebase} and exhibits three periods now. Results of the frequency analysis are given in Table \ref{tablefreq} (bottom). 
As in the two-planet resonant case, the long period of the TTVs signal corresponds to the precession 
rate of the arguments of the pericenters. The other two periods are related to $\theta_1$ and $\theta_2$, 
the libration angles of the 2:1 mean-motion resonance between the two inner planets, and $\theta_3$ 
and $\theta_4$, the libration angles of the 2:1 mean-motion resonance between the two outer planets. 
These two periods are both easily detected in the time evolution of the Laplace resonant 
argument $\phi=\lambda_1-3\lambda_2+2\lambda_3$ (see Figure \ref{figlgterme_casdebase} top right).

The three periods in the TTVs signal are a signature of the multiple resonance, it is not due 
to the fact that the system consists of three planets. To ilustrate this, we compare in Figure \ref{fig_dege} the TTVs of the same three-body system as before and another where the outer planet has been removed from the Laplace configuration.

\begin{figure}
 \centering
 \rotatebox{270}{\includegraphics[width=6cm]{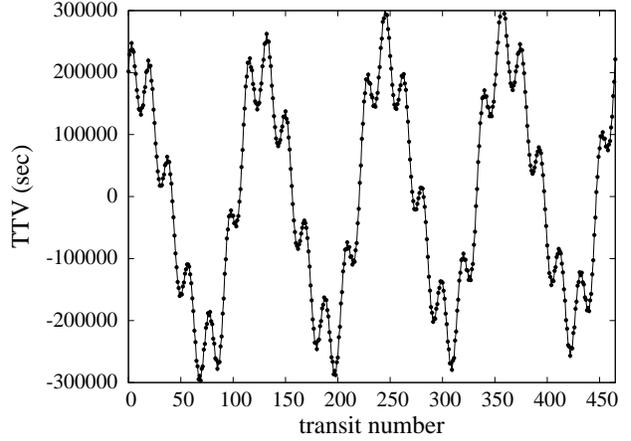}}
 \caption{TTVs of the inner planet of the Laplace-resonant three-planet Gliese 876 system. The parameters are issued from Table 3 of \citet{Riv10}.}
 \label{fig876}
 \end{figure}

 \begin{figure*}
 \centering
 \hspace{0.1cm}\rotatebox{270}{\includegraphics[height=8.5cm]{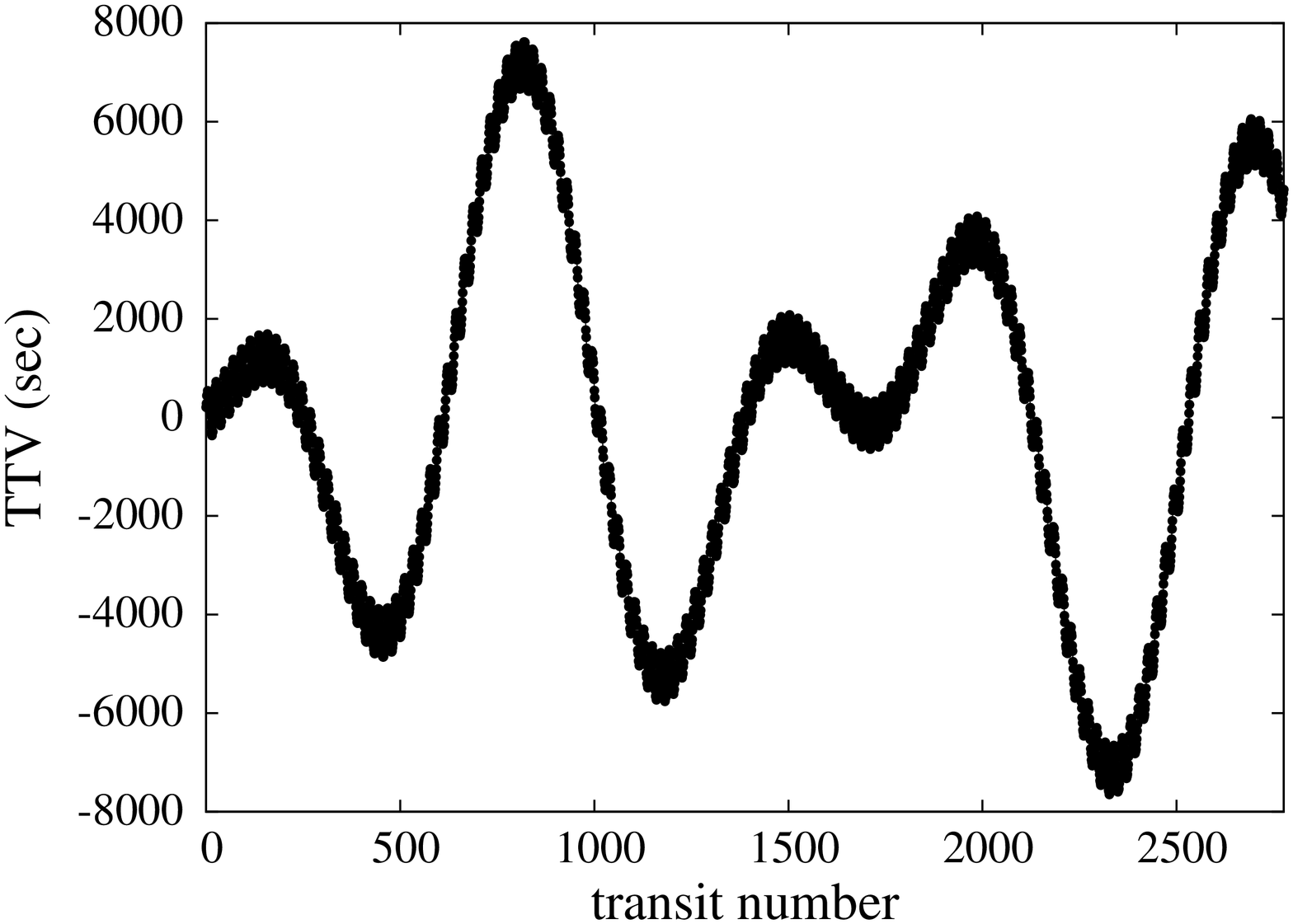}}
 \hspace{-0.1cm}\rotatebox{270}{\includegraphics[height=8.5cm]{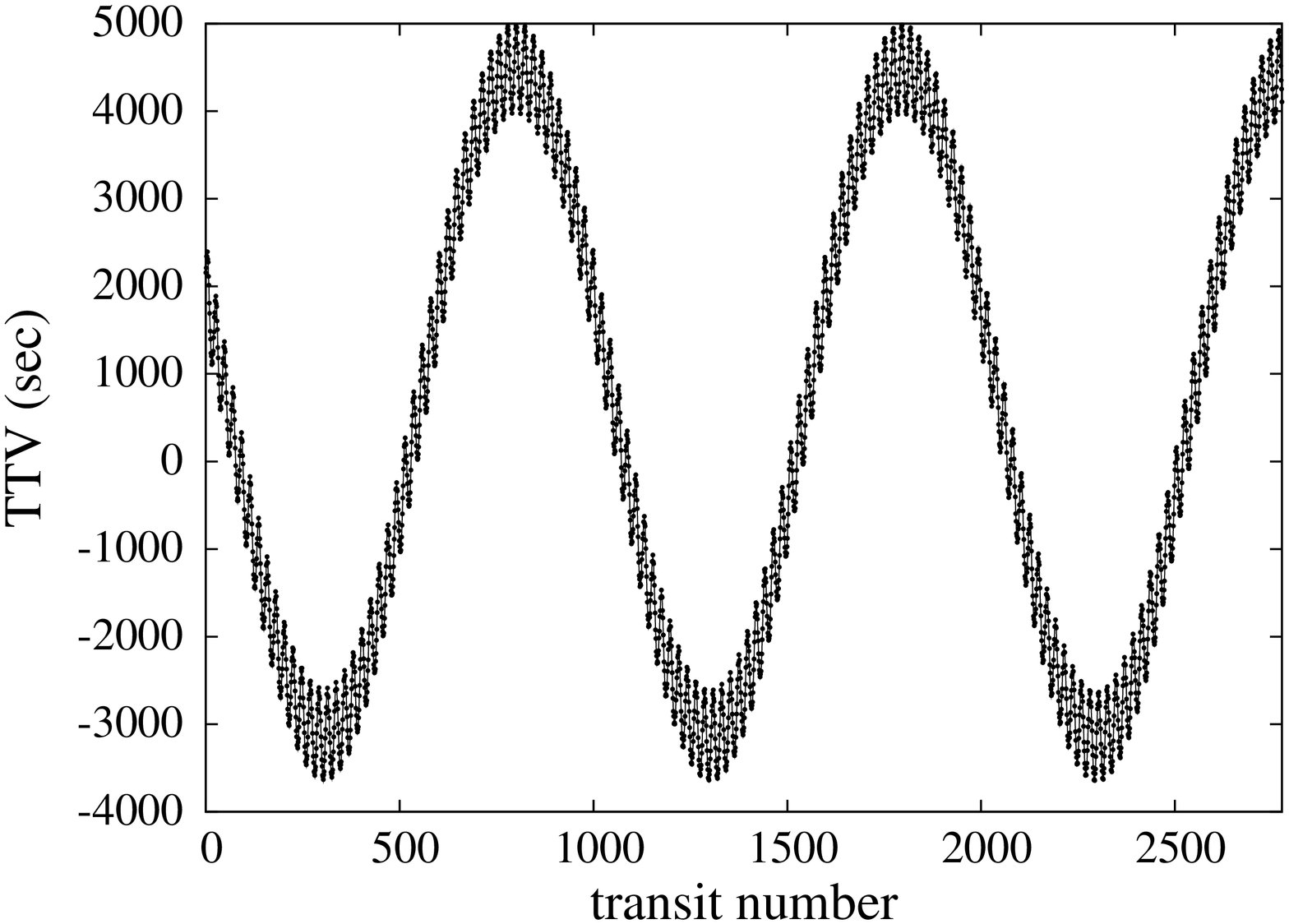}}
  \vspace{-1.8cm}
\rotatebox{270}{\includegraphics[width=5.9cm]{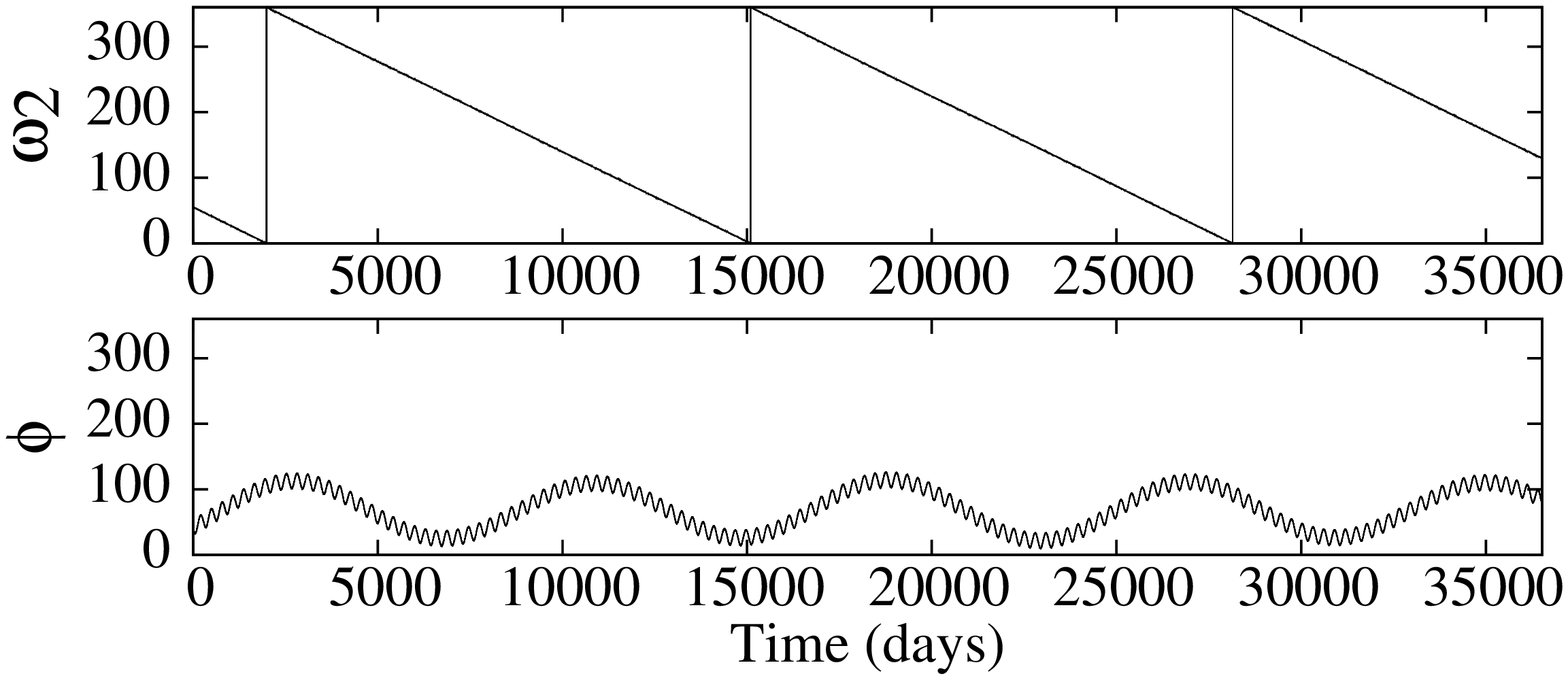}}
 \rotatebox{270}{\includegraphics[width=5.9cm]{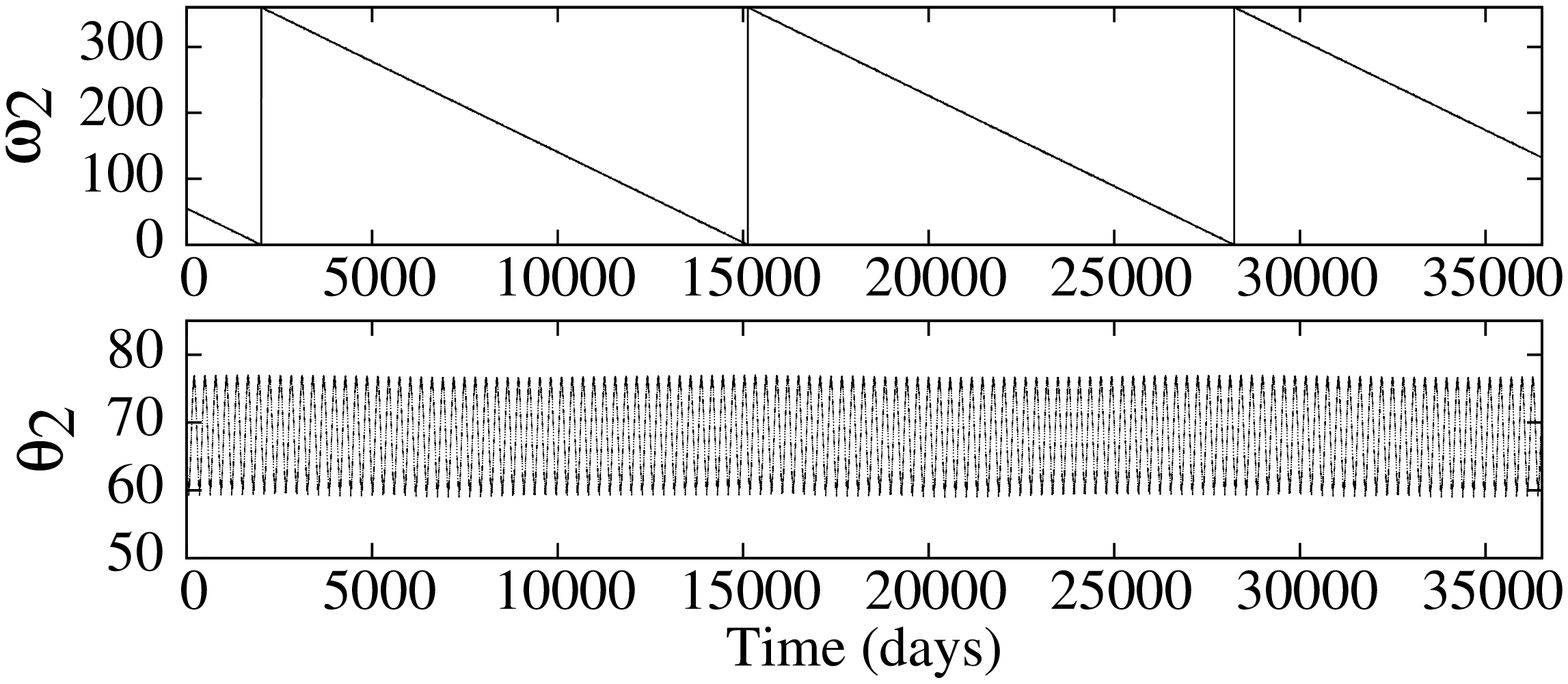}}
 
  \hspace{0.05cm}\rotatebox{270}{\includegraphics[height=8.5cm]{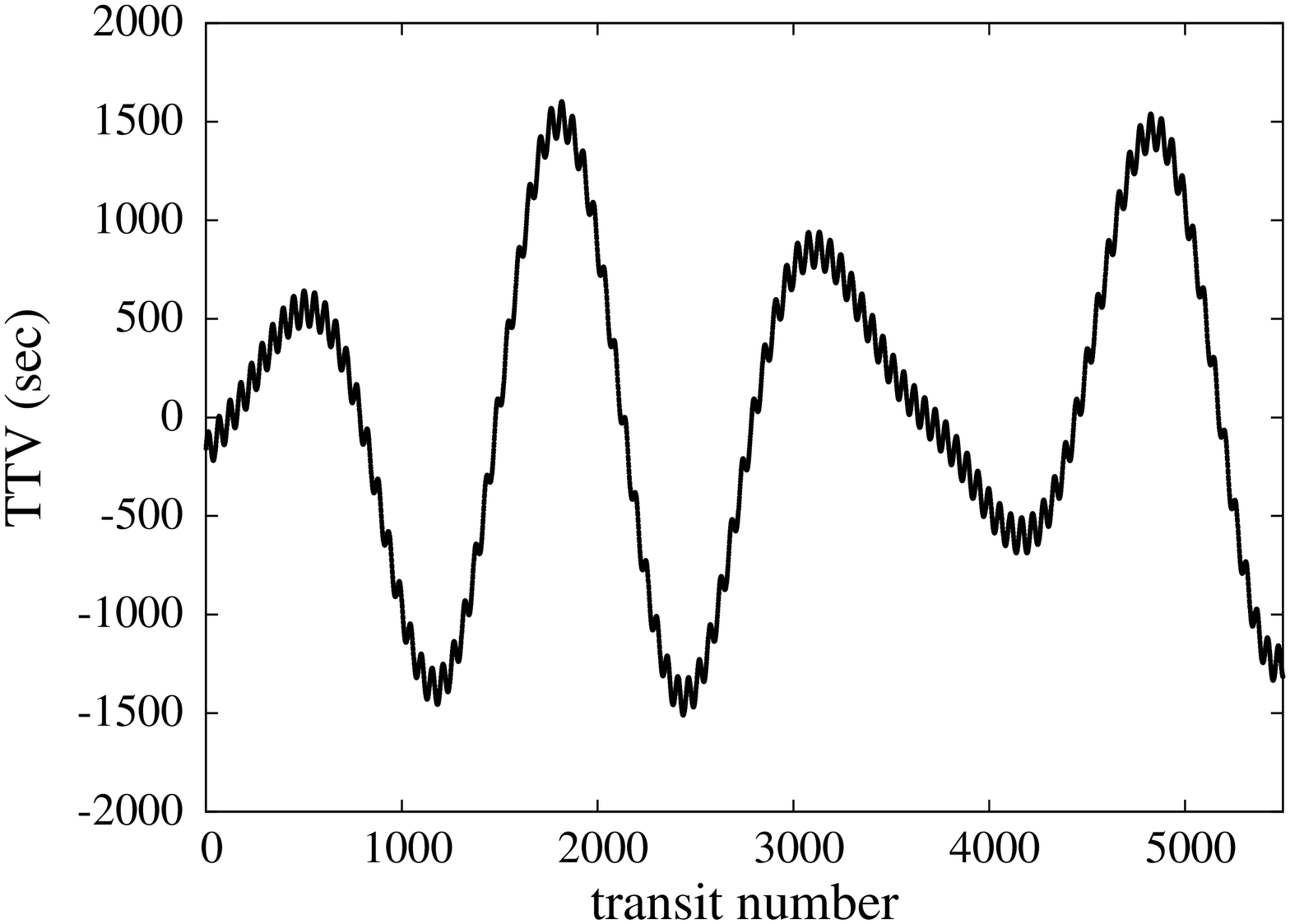}}
  \hspace{0.05cm}\rotatebox{270}{\includegraphics[height=8.5cm]{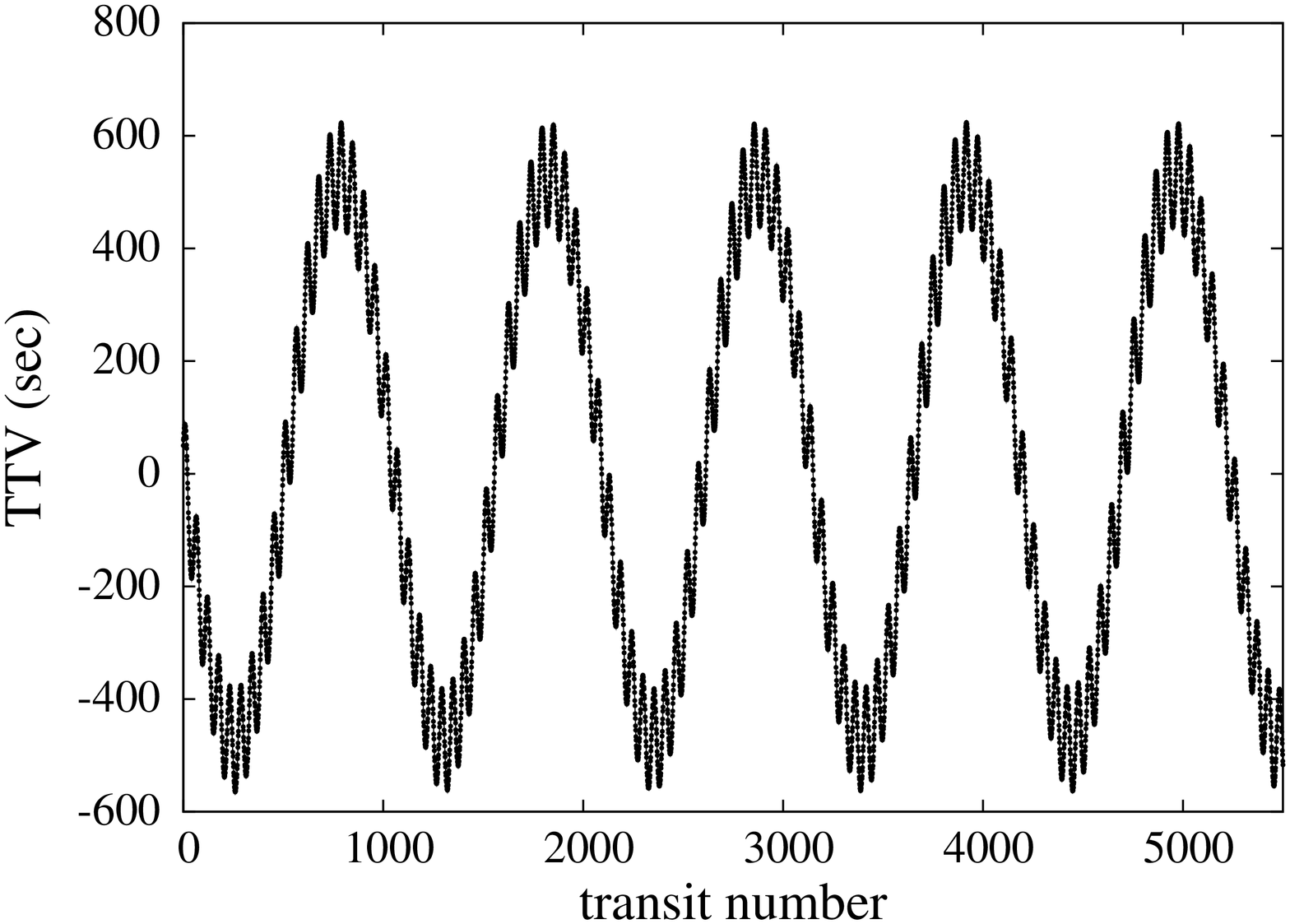}}
 \rotatebox{270}{\includegraphics[width=5.9cm]{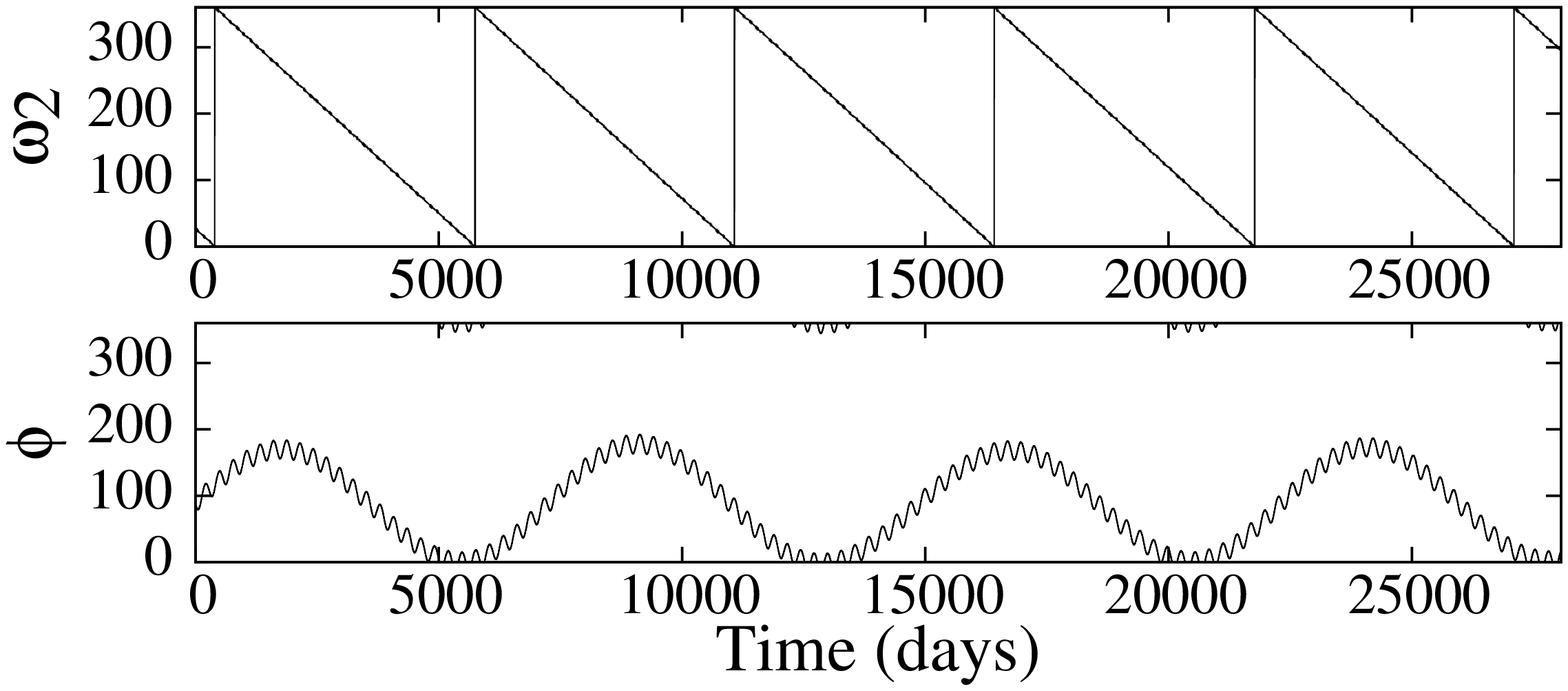}}
 \rotatebox{270}{\includegraphics[width=6.05cm]{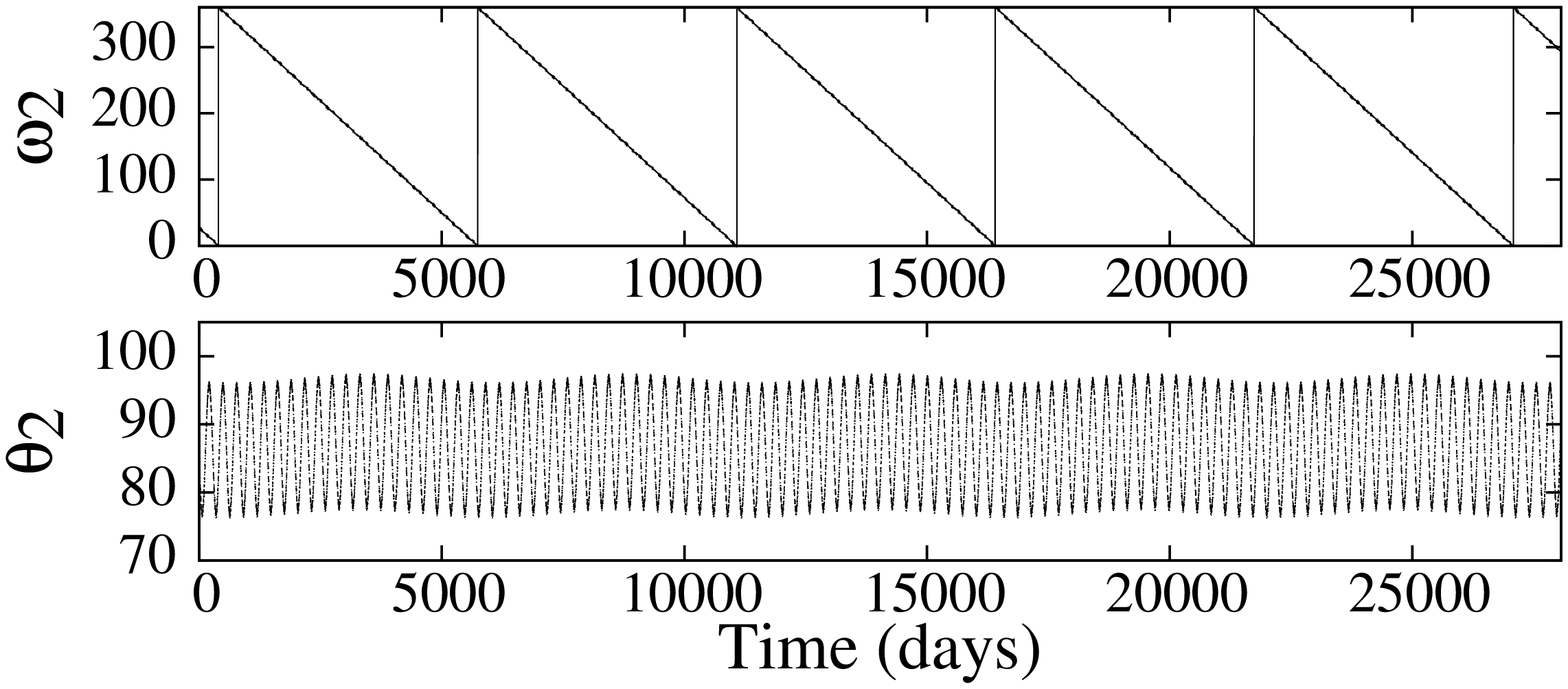}}

\vspace{-1.8cm} 
\caption{Left: TTVs and dynamical evolution of Laplace-resonant systems consisting of a giant 
planet and two terrestrial planets, on a time interval of $\sim 100$ years : a giant of $1 M_J$ with $P=13.1$ days + two super-Earths of $10 M_E$ on the top, a hot Jupiter of $0.5 M_J$ with $P=5.1$ days + two Earth-like planets of $2 M_E$ at the bottom. Right: Comparison with the systems formed by the two inner resonant planets only.}
 \label{fig2terres}
 \end{figure*}

 \begin{figure*}
 \centering
{ \rotatebox{270}{\includegraphics[width=4.cm]{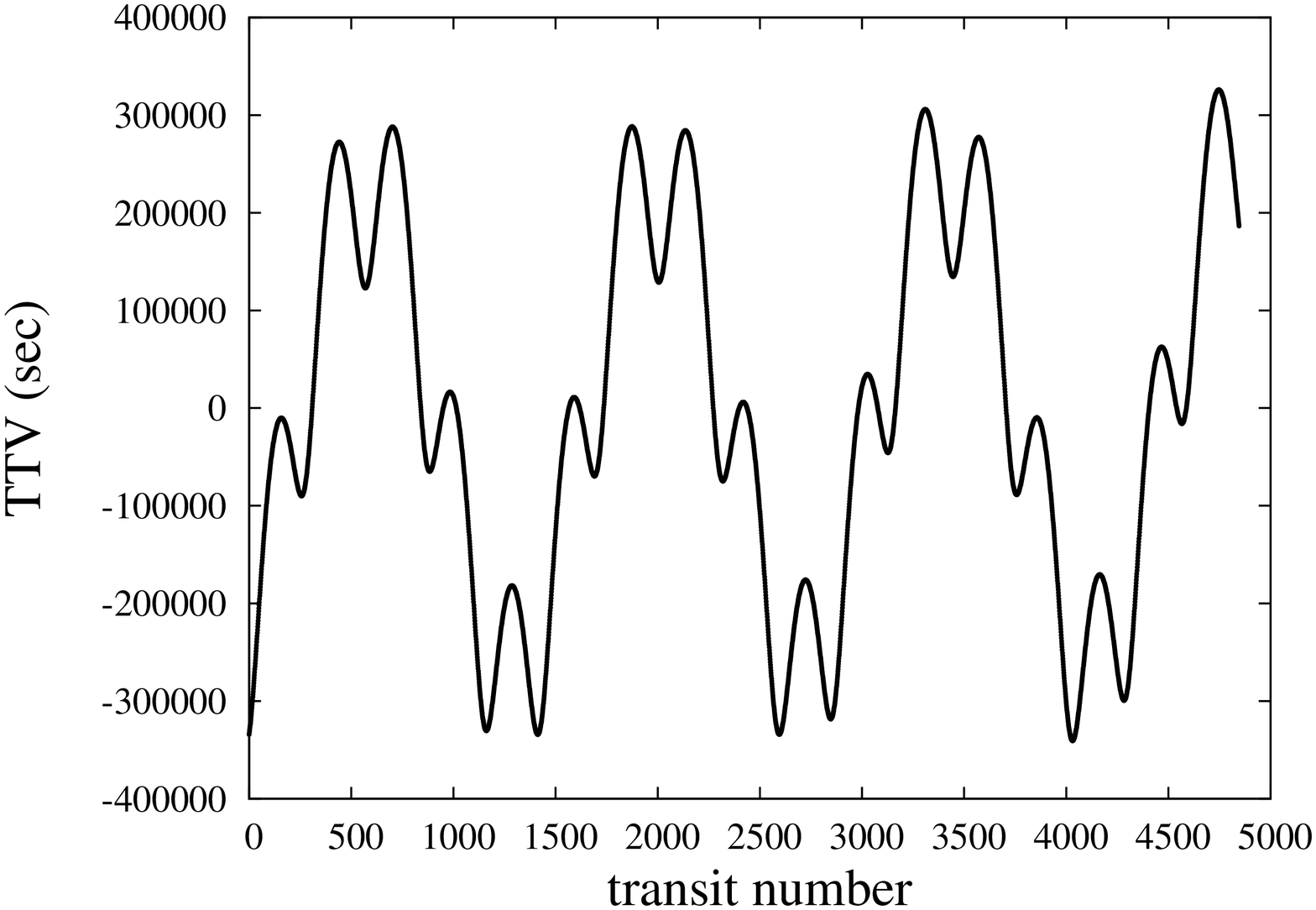}}
 \rotatebox{270}{\includegraphics[width=4.cm]{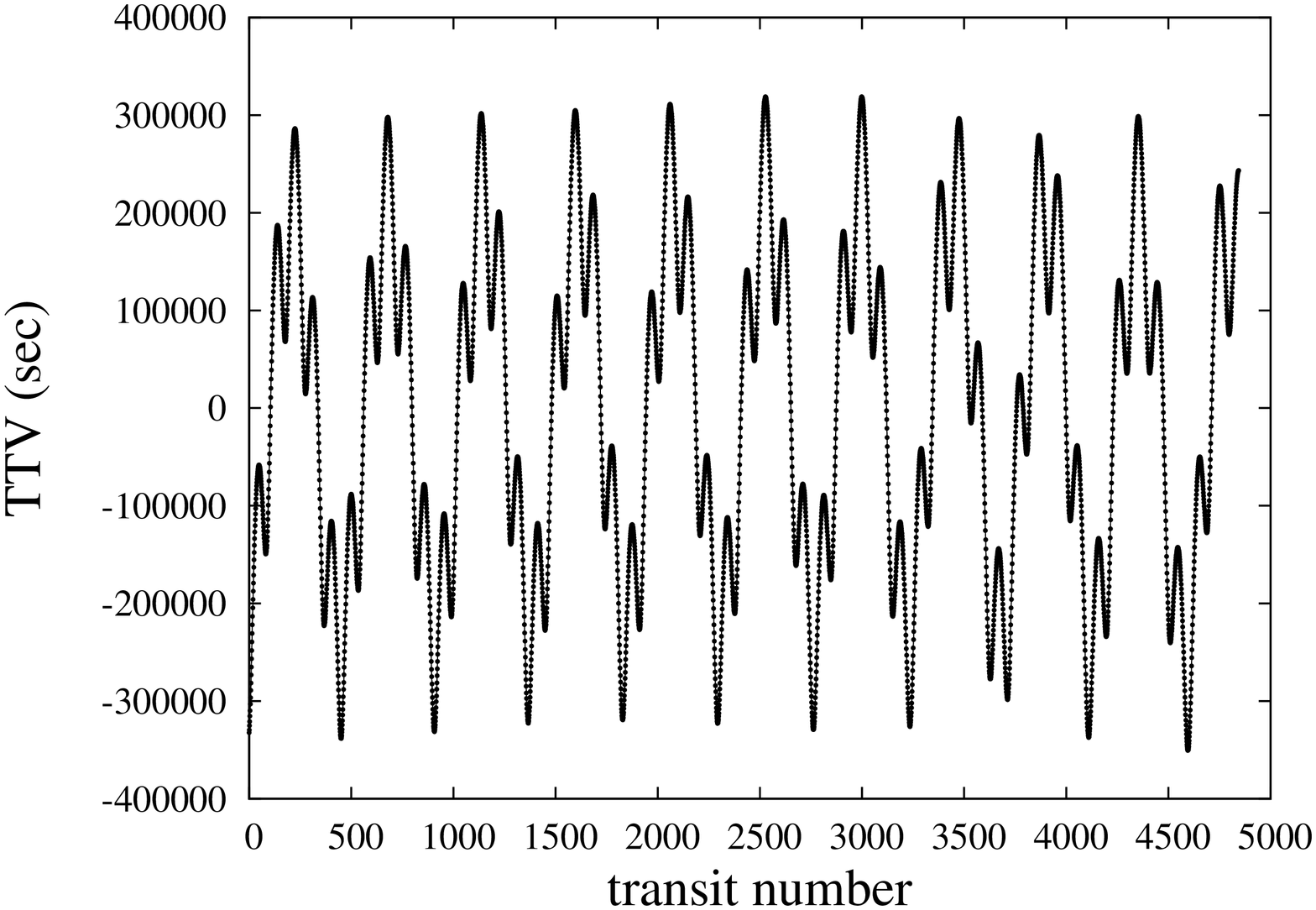}}
 \rotatebox{270}{\includegraphics[width=4.cm]{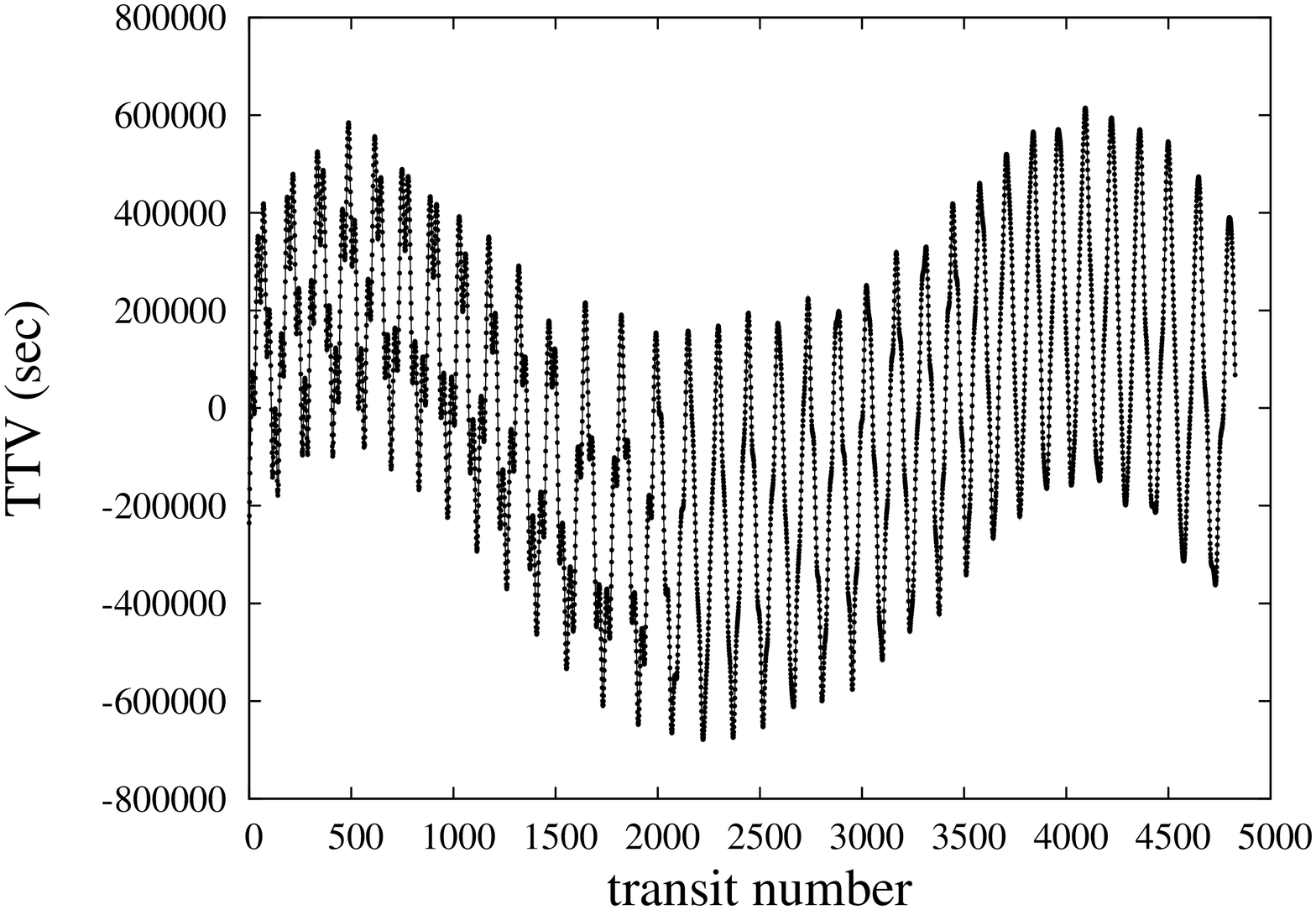}}
\rotatebox{270}{\includegraphics[width=3.8cm]{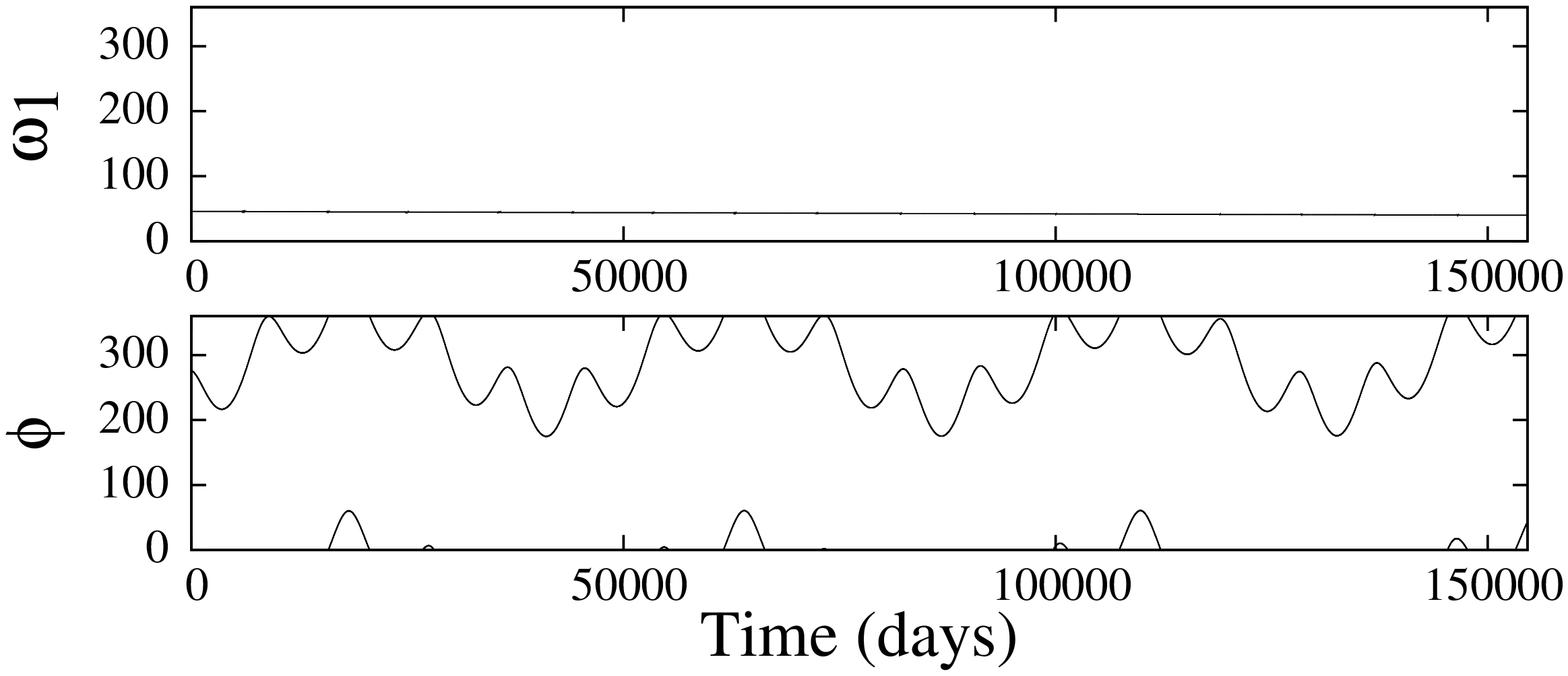}}
 \hspace{0.3cm}\rotatebox{270}{\includegraphics[width=3.8cm]{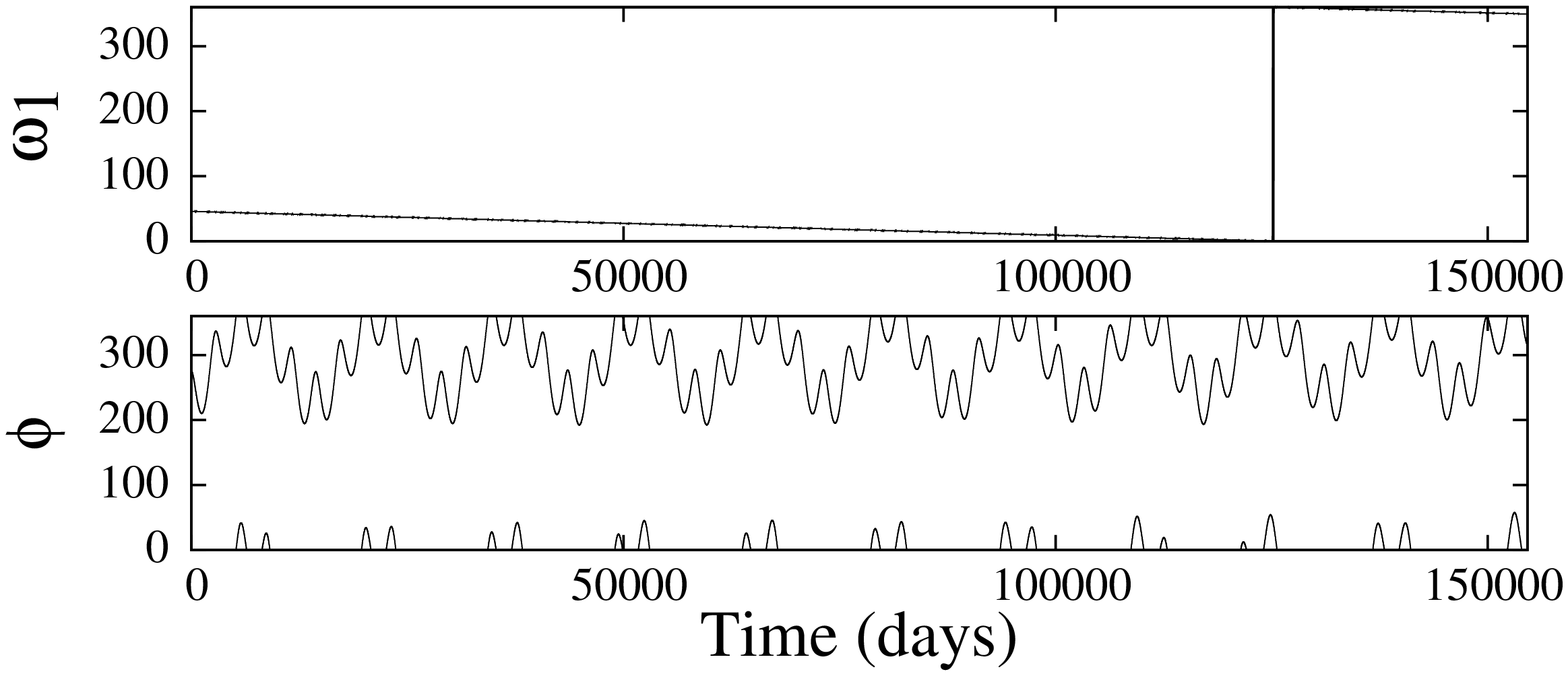}}
 \hspace{0.3cm}\rotatebox{270}{\includegraphics[width=3.8cm]{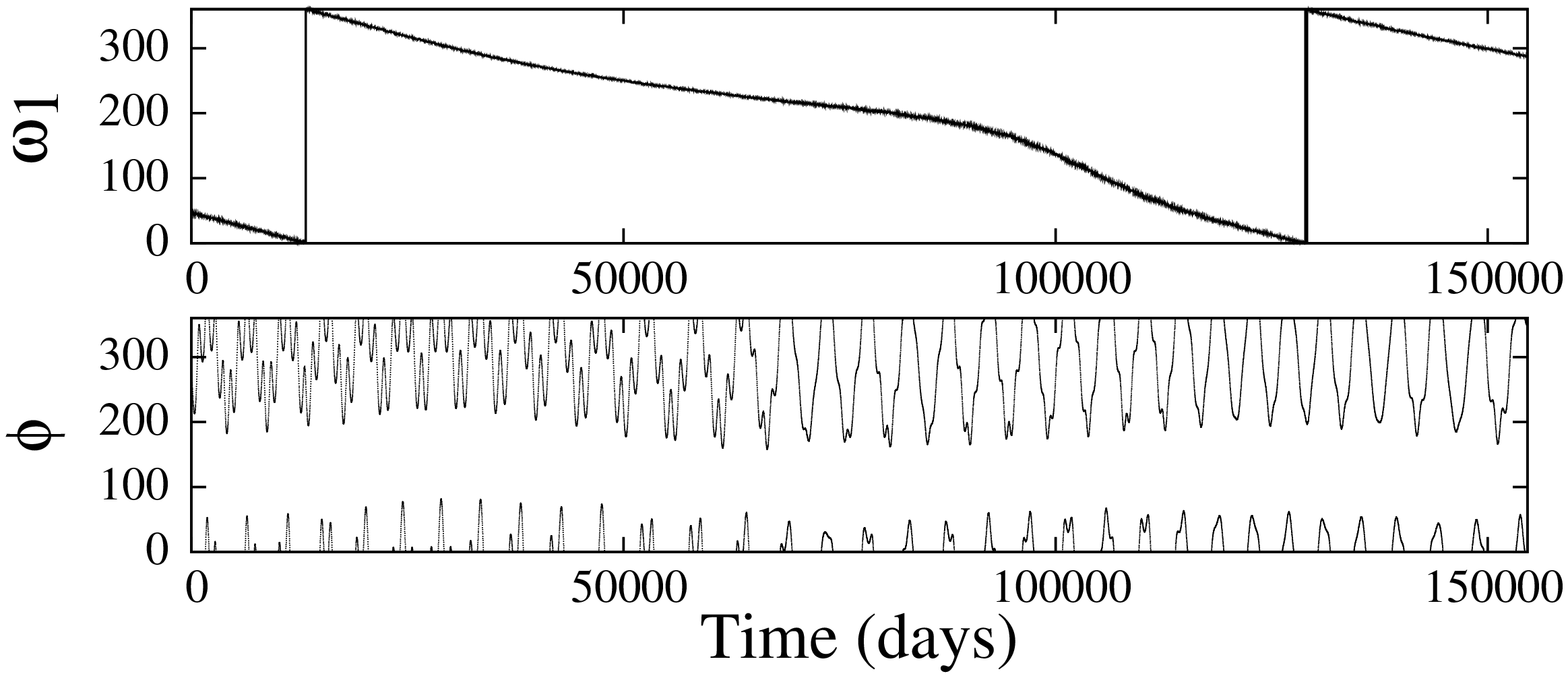}}
  \hspace{-0.4cm}}
\vspace{-1.5cm}
 \caption{TTVs and dynamical evolution on $\sim 400$ years for a system of 
 three Earth-like planets ($1.5 M_E$, left), three super-Earth planets ($15 M_E$, middle) 
 and three Jupiter-like planets ($1.5 M_J$, right). Periods of the planets are 31.9, 63.8 and
127.5 days, respectively. The same initial orbital elements have been used for the three systems.}
 \label{TTT}
 \end{figure*}

\section{On the possible detection of terrestrial planets}\label{secter}

We now investigate if the three-periods signature described in the previous section 
survives when considering one or more Earth-like planets in the Laplace resonance. Three different
configurations are examined, depending on the number of terrestrial planets in the system.

\subsection{One terrestrial and two giant planets}\label{sub1}
This case can be understood by studying Gliese 876 (\citet{Riv10}). 
It consists of two Jupiter-like planets ($0.7 M_J$ and $2.3 M_J$) 
and a Uranus-mass body ($14.6 M_E$, $M_E =$~Earth's mass) in the Laplace resonance
with orbital periods of $30.1$, $60.1$ and $124.3$ days, respectively. 
The TTVs of the inner planet are given in Figure~\ref{fig876}. 
Only two periods are detected in the curve, despite the Laplace configuration. This is confirmed with the frequency analysis. The same result holds for a Jupiter-Jupiter-Earth system.

Therefore, we conclude that the three-periods signature 
cannot be detected for a Jupiter-Jupiter-Earth Laplace-resonant configuration. 
This is a consequence of the weaker gravitational effect of the smallest planet on its giant companions.  Comparing Figures~\ref{fig876} and \ref{figlgterme_2pl}, we note that the TTVs curves look very similar. However, the presence of the third planet slightly modifies the period values and the amplitude of the signal. A deep analysis is needed to investigate if this difference can help to detect the third body. This is beyond the scope of this paper.

\subsection{Two terrestrial and one giant planets}\label{sub2}

This case is particularly interesting. We analyze a giant planet of $1 M_J$ in the Laplace resonance with two super-Earths of $10 M_E$, with orbital periods of $13.1$, $26.2$ and $52.5$ days, respectively. As in Section \ref{secTTVs}, the giant TTVs show three periods, see top left panel of Figure~\ref{fig2terres}. 
The first one is the shortest ($\sim 300$ days), arising from the 2:1 resonance between the two inner bodies.   
To identify the two other periods, the time variations of the Laplace-resonant angle $\phi$ and the pericenter argument $\omega_2$ are given. We remark that $\phi$ and the TTVs curve reach extrema at the same times. The period associated with the argument ${\omega_2}$ also appears in the TTVs, see top right panel of Figure~\ref{fig2terres}, where we consider, for comparison, the reduced system of the two inner planets only (in the 2:1 resonance).

As a result, the existence of two super-Earth companions of a giant planet in a Laplace-resonant configuration can be inferred from the TTVs of the giant. The amplitude of the TTVs signal is high ($\sim 2.2$ hours), and the time necessary to identify the three periods using frequency analysis (criterion : twice the longer period of the signal) is approximately $2000$ transits, equivalent to $\sim 70$ years in that case.

Another example is given at the bottom left panel of Figure~\ref{fig2terres}, with a 
system of a hot Jupiter of $0.5 M_J$ ($P=~5.1$~days) and two terrestrial planets of $2 M_E$ ($P=10.2$ and $20.5$ days, respectively). Since the terrestrial masses are smaller than in the previous case, $\sim 3000$ transits ($\sim 40$ years) are needed to detect the three periods by frequency analysis.

\subsection{Three terrestrial planets}\label{sub3}

Finally, let us consider a Laplace-resonant system of three terrestrial planets ($1.5 M_E$) with periods of $31.9$, $63.8$ and $127.5$ days, respectively. The TTVs curve of the inner planet is shown in Figure \ref{TTT} (left panel). As expected, the amplitude of the TTVs signal is extremely high ($\sim 3.5$ days), but only two periods are detectable. The long period associated with the argument ${\omega}$ is 
longer than the time-span considered, and thus visible by increasing the mass values up to $0.5 M_J$ (middle and right panels). Indeed the smaller the masses of the planets, the longer
the precession rate of the arguments of the pericenters.

Therefore, for a system of three Earth-like planets in a multiple resonance, 
the secular period cannot be detected in the TTVs in a time-span of $\sim 100$ years. 
Only the two periods associated with the Laplace angle can be deduced, the longer being, from our example, $\sim 1500$ transits ($\sim 120$ years) for $1.5 M_T$ planets. Nevertheless, the double-frequency TTVs signal is here the signature of a double 2:1 resonance between the three Earth-like bodies, with amplitudes of $\sim~42$ and $83$ hours. This can be retrieved from two-body resonance calculations (e.g. Section 6 of \citet{Ago05}).

 \begin{figure}
 \hspace{0.4cm}
 \rotatebox{270}{\includegraphics[width=5.8cm]{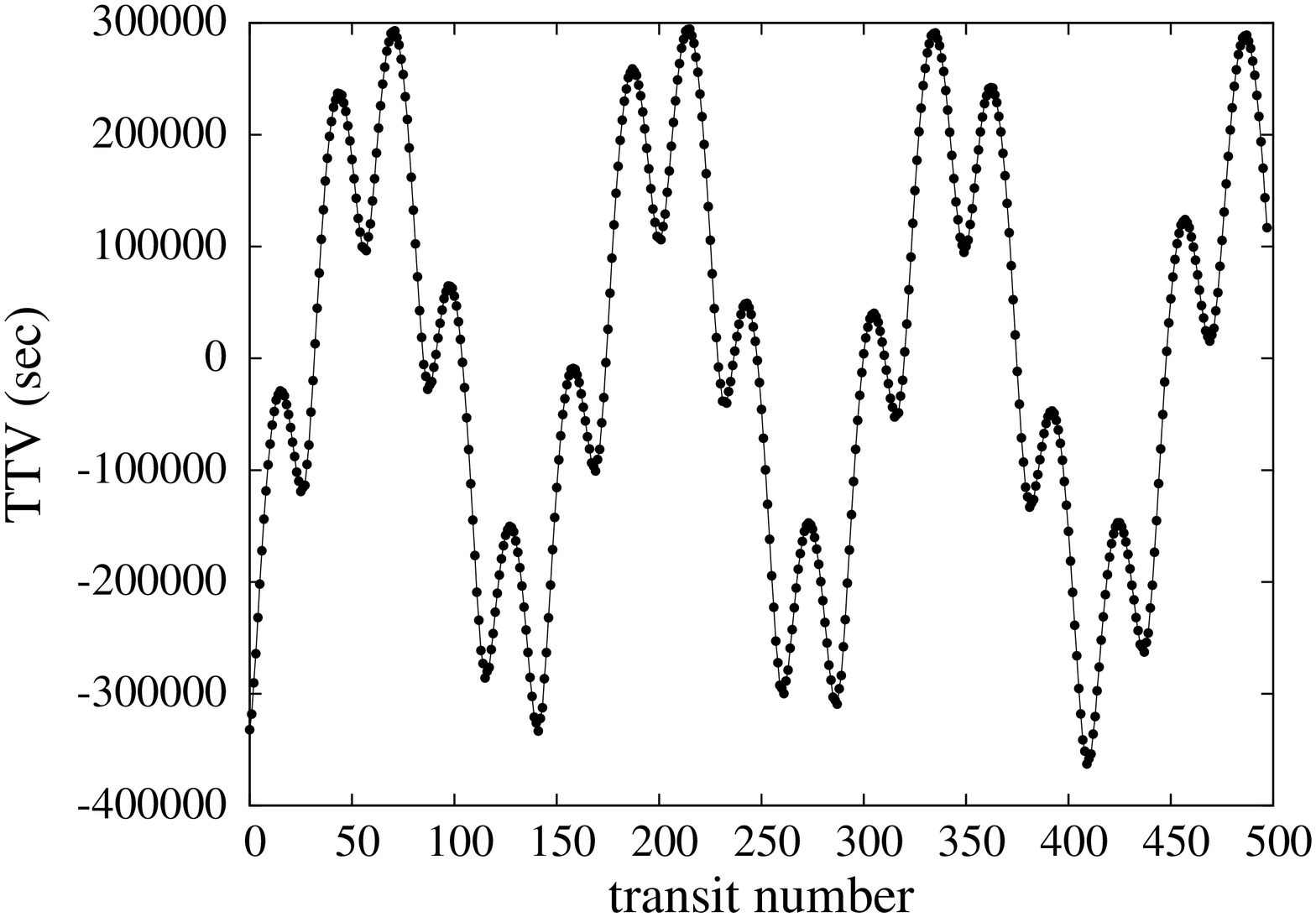}}
 \rotatebox{270}{\includegraphics[width=6.2cm]{TTV_Gl876_2pl}}
 \caption{Double period TTVs for different system configurations on $\sim 40$~years.
 Top panel: the three Jupiter-like planets given in Figure~\ref{TTT}. Bottom panel:
 the Gliese 876 two-planet system shown in Figure~\ref{figlgterme_2pl}. In each case the 
 transiting giant planet is $\sim 0.5 M_J$ with an orbital period of $\sim 30$ days.}

 \label{short}
 \end{figure}

\section{Discussion of the results}\label{secdisc}

\subsection{Limitations due to observation time-span}
We have shown in the previous section that at least several decades are necessary to practically 
identify Laplace-resonant three-planet systems containing terrestrial bodies. On a given short observation time-span, only two periods can usually be detected in 
the TTVs of a Laplace-resonant system, thus leading to a curve very similar to the one
of two planets in resonance. This is illustrated in Figure~\ref{short}: the long-period resonant argument of a three-Jupiter system matches the pericenter precession of a two-body configuration. This example suggests possible degeneracy between two- and three-planet resonant configurations on short observation time-spans, and this question should be investigated in more details in the future.

Therefore, long-term TTVs observations are necessary to constrain dynamically 
multiple-planet resonant systems, especially if no additional data such as radial 
velocity data are available. Let us note that, even if this work identifies
the three periods in the TTVs signal as a signature of a resonant three-body configuration, 
the same three periods could correspond to different parametrizations of the system, as the degeneracy 
in the characterization of non-transiting planets pointed out for a two-body resonance (see e.g. \citealt{Bou12}). 
Furthermore, only coplanar systems have been investigated here. The detection
 of inclined Laplace-resonant systems should also be adressed, as the
 possible formation of (highly) inclined two- and three-body resonant systems 
 has recently been demonstrated (e.g. \citealt{tho03,Lib09,Lib11b}).

\subsection{Kepler systems}

At the time of writing, three confirmed Kepler systems consist of three planets: 
Kepler-9 (\citealt{Hol10}), Kepler-18 (\citealt{Coc11}) and Kepler-30 (\citealt{Fab12}). 
The TTVs of Kepler-9b and Kepler-9c present the characteristic signature of a 2:1 mean-motion resonance. These two giant planets also have an inner non-resonant super-Earth companion. Kepler-18 exhibits a similar dynamics, with two Neptune-mass planets in a 2:1 mean-motion resonance confirmed by TTVs and an inner non-resonant super-Earth planet. As argued in Section \ref{sub1}, such systems could host a terrestrial Laplace-resonant planet, but this would require further investigations. 
Kepler-30 is an interesting three-planet system with significant TTVs. While the 
Neptune-size Kepler-30b lies interior to a 2:1 mean-motion resonance with the 
Jupiter-size Kepler-30c, an additional Jupiter-size planet orbits the star with 
a period that roughly corresponds to $5$ times the one of Kepler-30b and $2.5$ 
times the one of Kepler-30c. The possible proximity to such a three-planet resonance 
could explain the difficult interpretation of the TTVs on the current short observation time. 
A longer time-span will help to identify the possible signature of three periods in 
the signal - if they exist. Let us note that most recent Kepler systems (e.g. Kepler-51, Kepler-53 and Kepler-60) show potential multibody resonances (see \citealt{Ste12}). Furthermore 
a lot of KOIs are close to 
multiple period ratio commensurabilities
For instance, KOI-500 (\citealt{Lis11}) is particularly interesting with possibly two 
three-body resonances. However, a study of the inner TTVs curve only is not sufficient 
for the dynamics of a five-body system.

\section{Conclusion}\label{ccl}

We studied the possible detection of Laplace-resonant three-planet systems 
on a maximal time-span of $\sim 100$ years, assuming that only the inner planet transits the star.
We have shown that a three-body resonant configuration has a specific TTVs shape: three periods are present, the long one associated with the precession of the arguments of the pericenters, and the other two with the time evolution of the Laplace-resonant angle. Our criterion for detecting a signal is that the frequency is sampled completely, not that the amplitude of the Fourier component is large.      

The three periods are detectable for a giant planet perturbed by two terrestrial bodies. 
A system of three Earth-like planets can be constrained from the inner planet TTVs, even if only 
two periods are visible. We discussed the number of transits necessary to secure a detection. 
We pointed out that two and three-body resonant configurations could be mistaken on a short observation time, as the long-period resonant argument of a three-Jupiter system can be confused with the pericenter precession of a two-body configuration.

Many Kepler extrasolar systems (and KOIs) reveal to be in a closely packed configuration 
(\citet{Lis11}), and multiple-planet resonances can act as a phase-protection mechanism which insures long-term stability. Formation scenarios (during the gas-disk phase) can also favor
multiple resonant systems. We emphasize the importance of TTVs long-term observing programs for 
the detection of such systems in the future and the contraints it will provide on the 
formation history of planetary systems.

\section*{Acknowledgement}
We thank the anonymous referee for his/her useful suggestions. The work of A-S L is supported by an F.R.S.-FNRS Postdoctoral Research Fellowship. A-S L thanks the people at Lille Observatory for their hospitality during her stay.

\bibliographystyle{}

\end{document}